\documentclass[twocolumn]{article}
\usepackage{parskip}
\usepackage{amssymb}
\usepackage{amsmath}
\usepackage{fullpage}
\usepackage{authblk}
\usepackage{blindtext}
\usepackage{siunitx}
\usepackage{float}
\usepackage{caption}
\usepackage[superscript,nomove]{cite}
\usepackage{hyperref}
\usepackage[normalem]{ulem}
\usepackage{graphicx}
\usepackage{cancel}
\usepackage{pifont}
\usepackage{diagbox}
\usepackage[percent]{overpic}
\usepackage{tikz}

\makeatletter \renewcommand{\@citess}[1]{\textsuperscript{[#1]}} \makeatother

\makeatletter
\def\@firstoftwo@second#1#2{%
  \def\temp##1.##2\@nil{##2}%
   \temp#1\@nil}
\newcommand\sref[1]{%
   (A.\expandafter\@setref\csname r@#1\endcsname\@firstoftwo@second{#1})%
}

\makeatother

\title{A generalized and adaptable tensor-contraction-based cluster expansion formalism for multicomponent solids}
\author[1]{\small Jacob Jeffries \thanks{jwjeffr@g.clemson.edu}}
\author[1]{Bochuan Sun}
\author[1,2]{Enrique Martinez \thanks{enrique@clemson.edu}}
\affil[1]{Department of Materials Science and Engineering, Clemson University, Clemson, SC 29634, USA}
\affil[2]{Department of Mechanical Engineering, Clemson University, Clemson, SC 29634, USA}
\date{\small \today}

\renewenvironment{abstract}
 {\quotation\small\noindent\rule{\linewidth}{.5pt}\par\smallskip
  {\centering\bfseries\abstractname\par}\medskip}
 {\par\noindent\rule{\linewidth}{.5pt}\endquotation}

\begin{document}

\twocolumn[
  \begin{@twocolumnfalse}
  \maketitle
    \begin{abstract}
        Density functional theory (DFT)-based simulations of materials have first-principles accuracy, but are very computationally expensive. For simulating various properties of multi-component alloys, the cluster expansion (CE) technique has served as the standard workaround to improve computational efficiency. However, the standard CE technique is difficult to extend to exotic and/or low-symmetry lattices, often implemented via iteration over particular cluster types, which must be enumerated per lattice structure. In this work, we introduce the tensor cluster expansion (TCE), implemented in the open-source code tce-lib, which maps correlation functions to mixed tensor contractions, eliminating the need to iterate over cluster types and additionally making the calculation of correlation functions well-suited for massively parallel architectures like GPUs. We show that local interaction energies are an immediate consequence of the TCE formalism, yielding nearly $\mathcal{O}(1)$ energy difference calculations. We then use this formalism to fit CE models for the TaW and CoNiCrFeMn systems, and use these models to respectively compute the enthalpy of mixing curve and Cowley short-range order parameters, showing excellent agreement with ground truth data.
    \end{abstract}
  \vspace{0.5cm}
  \end{@twocolumnfalse}
]

\section{Introduction}

Atomistic simulations have become an essential tool for understanding and predicting the properties and behavior of various systems at the atomic scale, providing insight into phenomena that are often challenging to access experimentally, including (but not limited to) phase stability\cite{mishin2005phase, cuesta2013structure, abraham1981phases, sepliarsky2011first, kaczmarski2005phase}, mechanical properties\cite{bu2009atomistic, rodriguez2014atomistic, coluci2007atomistic, wang2008atomistic, zhang2021prediction}, and effects of ordering phenomena\cite{sheng2006atomic, ferrari2023simulating, woodgate2023short, jeffries2025prediction}. Among the methods for driving an atomistic simulation, density functional theory (DFT)\cite{hohenberg1964inhomogeneous, kohn1965self} offers first-principles level accuracy of material properties, enabling highly reliable calculations of electronic structure and the resulting energies and forces, yielding accurate predictions of various material properties, including phase boundaries \cite{ceperley1980ground}, lattice parameters\cite{jain10materials}, and more. However, DFT is very computationally expensive, and is infamously restricted to relatively small systems, usually on the order of $10^2$-$10^3$ atoms\cite{payne1992iterative, bowler2012methods, martin2020electronic}. Classical interatomic potentials, such as embedded-atom method (EAM)\cite{daw1984embedded, daw1993embedded} and modified embedded-atom method (MEAM)\cite{baskes1992modified, lee2000second, lee2001second} potentials, are able to push this size envelope, but often suffer from limited transferability\cite{muser2023interatomic, rassoulinejad2016evaluation}, sensitivity to fitted parameters\cite{MAHMOOD2024115900}, and questionable predictive capabilities when compared to experimental data\cite{Choudhary2017, PhysRevB.59.3393, mendelev2008analysis, vella2015comparison}.

For multi-component solids, an alternative approach is the cluster expansion (CE) technique\cite{annurev:/content/journals/10.1146/annurev-matsci-070317-124443, FONTAINE199433}, which leverages a relatively small set of DFT calculations to train an effective Hamiltonian in terms of correlation functions on a static lattice:

\begin{equation}\label{eq:old-ce}
    \mathcal{H}_\text{eff}(\sigma) = \sum_\alpha m_\alpha J_\alpha \Pi_\alpha(\sigma)
\end{equation}

where $\alpha$ denotes cluster types that are unique under symmetry operations, $m_\alpha$ denotes the multiplicity of the $\alpha$ cluster type, i.e. the number of symmetrically equivalent clusters, $\Pi_\alpha(\sigma)$ denotes the correlation function of a configuration $\sigma$ for cluster type $\alpha$, $J_\alpha$, a learnable parameter, is the effective cluster interaction (ECI) of cluster $\alpha$, and $\mathcal{H}_\text{eff}$ is the effective Hamiltonian, or the approximate energy, of a configuration $\sigma$. This technique has been shown to be very effective at estimating thermodynamic properties of multi-component solids, including order-disorder transition temperatures and miscibility gaps\cite{PRXEnergy.3.042001}.

In popular CE codes, such as Alloy Theoretic Automation Toolkit (ATAT)\cite{atat}, Integrated Cluster Expansion Toolkit (ICET)\cite{icet}, and the Clusters Approach to Statistical Mechanics (CASM) toolset\cite{casm}, correlation functions are computed by iterating over symmetrically-equivalent clusters. This approach is simple, elegant, and flexible for handling differing cluster types on a conventional lattice, but sequentially processes many small, per-cluster loops with scattered memory access, and is therefore difficult to fully exploit the vectorization capabilities of massively parallel architectures such as graphic processing units (GPUs) and tensor processing units (TPUs). This approach additionally requires explicit enumeration of new cluster types for exotic and/or low-symmetry lattices.

In this work, we present a new CE formalism, named the tensor cluster expansion (TCE), which evaluates the necessary correlation functions entirely as mixed sparse-dense tensor contractions between precomputed topology tensors and one-hot-encoded configuration tensors. Because the TCE framework only depends upon precomputed topology tensors, which can be obtained for any periodic lattice, no new data enumeration is needed for exotic or low-symmetry lattices. Additionally, the TCE approach fully eliminates the need for per-cluster loops, instead replacing them with tensor contractions, enabling regular memory access patterns that are well suited for vectorization and execution on GPUs and/or TPUs. This formalism is then implemented in our open-source Tensor Cluster Expansion Library\cite{tce_lib} package, \verb|tce-lib| for short, written in Python and leveraging the \verb|sparse|\cite{abbasi2025pydatasparse} and \verb|opt-einsum|\cite{opt_einsum} packages for efficient and user-friendly sparse tensor contraction.

We then show that the computation of local interaction energies is a direct consequence of this formalism, making the efficient calculation of energy differences trivial. We additionally test this approach with two numerical experiments: computing the heat of mixing curve of the $\text{Ta}_{1-x}\text{W}_{x}$ alloy using DFT data, and the equilibrium short range order of the equiatomic CoNiCrFeMn high entropy alloy using a MEAM potential.

\section{Model}

\subsection{Effective Hamiltonian}

In this work, we define an effective Hamiltonian $\mathcal{H}_\text{eff}$ of an atomic configuration $\mathbf{X}$ as a decomposition of many-body interactions:

\begin{equation}
    \begin{aligned}
        \mathcal{H}_\text{eff}(\mathbf{X}) &= \text{two-body interactions}\\
                               &+ \text{three-body interactions}\\
                               &+ \cdots
    \end{aligned}
\end{equation}

In terms of the number of $n$'th order two-body $\alpha$-$\beta$ interactions $N_{\alpha\beta}^{(n)}$ and the number of $n$'th order $\alpha$-$\beta$-$\gamma$ interactions $M_{\alpha\beta\gamma}^{(n)}$, the effective Hamiltonian can be written in terms of learnable interaction energies:

\begin{equation}\label{eq:effective-hamiltonian}
    \begin{aligned}
        \mathcal{H}_\text{eff}(\mathbf{X}) &= \frac{1}{2!}\,\varepsilon^{(n)}_{\alpha\beta}\, N^{(n)}_{\alpha\beta} + \frac{1}{3!}\,\zeta^{(n)}_{\alpha\beta\gamma}\, M^{(n)}_{\alpha\beta\gamma} + \cdots \\
        &=\frac{1}{2!}\hspace{-7.5pt}\begin{tikzpicture}[baseline={(0,0)}]
            \node[circle, minimum size=15pt, inner sep=0, fill=red!30, draw=black] (E2) at (0, 0) {$\varepsilon$};
            \node[circle, minimum size=15pt, inner sep=0, fill=blue!30, draw=black] (N2) at (1.5, 0) {$N$};

            \draw (E2.north east) to[out=40, in=140] node[midway, above] {$\alpha$} (N2.north west);
            \draw (E2.south east) to[out=320, in=220] node[midway, below] {$\beta$} (N2.south west);
            \draw (E2.south) .. controls +(-0.6,-1.0) and +(0.6,-1.0) .. node[midway,below]{$n$} (N2.south);
        \end{tikzpicture}\hspace{-7.5pt} + 
        \frac{1}{3!}\hspace{-7.5pt}\begin{tikzpicture}[baseline={(0,0)}]
            \node[circle, minimum size=15pt, inner sep=0, fill=green!30, draw=black] (E3) at (0, 0) {$\zeta$};
            \node[circle, minimum size=15pt, inner sep=0, fill=yellow!30, draw=black] (N3) at (1.5, 0) {$M$};

            \draw (E3.north east) to[out=40, in=140] node[midway, above] {$\alpha$} (N3.north west);
            \draw (E3.east) to node[midway, above, yshift=-2.5pt] {$\beta$} (N3.west);
            \draw (E3.south east) to[out=320, in=220] node[midway, below] {$\gamma$} (N3.south west);
            \draw (E3.south) .. controls +(-0.6,-1.0) and +(0.6,-1.0) .. node[midway,below]{$n$} (N3.south);
        \end{tikzpicture}\hspace{-7.5pt} + \cdots
    \end{aligned}
\end{equation}

where we have adopted the Einstein summation convention, and Greek letters denote chemical types. Additionally, we have used the Penrose graphical notation to visually depict the contractions, where each node represents a tensor, each free edge represents a free index, and each remaining edge represents a contraction along a given axis\cite{penrose1971applications, taylor2024introduction}. Here, $\varepsilon_{\alpha\beta}^{(n)}$ can be interpreted as the interaction energy of an $\alpha$-$\beta$ $n$'th order pair. Similarly, $\zeta_{\alpha\beta\gamma}^{(n)}$ denotes the energy of an $\alpha$-$\beta$-$\gamma$ $n$'th order triplet. Here, we use a one-hot encoded tensor of the configuration $\mathbf{X}$:

\begin{equation}
    X_{i\alpha} = \begin{cases}
        1 & \text{site $i$ is occupied by type $\alpha$} \\
        0 & \text{else}
    \end{cases}
\end{equation}

where Latin indices denote lattice sites. This encoding equips each lattice site with a Boolean vector $\mathbf{X}_i$, with length equal to the number of chemical species. For example, for a ternary FeNiCr alloy, and if a lattice site $i$ is occupied by an Fe atom, then $\mathbf{X}_i = (X_{i, \text{Fe}}, X_{i, \text{Ni}}, X_{i, \text{Cr}}) = (1, 0, 0)$. 

Additionally, this encoding allows us to decompose $N_{\alpha\beta}^{(n)}$ and $M_{\alpha\beta\gamma}^{(n)}$ in terms of the systems topology. Including up to three-body interactions, the topology of the system is fully defined by the topology tensors $A_{ij}^{(n)}$ and $B_{ijk}^{(n)}$, where two-body interactions are encoded by the first:

\begin{equation}
    A_{ij}^{(n)} = \begin{cases}
        1 & \text{sites $i$ and $j$ are $n$'th neighbors} \\
        0 & \text{else}
    \end{cases}
\end{equation}

and three-body interactions are encoded by the latter:

\begin{equation}
    B_{ijk}^{(n)} = \begin{cases}
        1 & \text{sites $i$, $j$, and $k$ are an $n$'th order triplet} \\
        0 & \text{else}
    \end{cases}
\end{equation}

yielding:

\begin{equation}
    \begin{aligned}
        N_{\alpha\beta}^{(n)} &= A_{ij}^{(n)}X_{i\alpha}X_{j\beta}
        =\begin{tikzpicture}[baseline={(0,0)}]
          \node[circle, minimum size=15pt, inner sep=0, fill=violet!60, draw=black] (A) at (1.25,0) {$A$};
          \node[circle, minimum size=15pt, inner sep=0, fill=orange!60, draw=black] (X1) at (0, 0.5) {$X$};
          \node[circle, minimum size=15pt, inner sep=0, fill=orange!60, draw=black] (X2) at (0, -0.5) {$X$};

          \draw (A.north west) to node[midway, above]{$i$} (X1.east);
          \draw (A.south west) to node[midway, below] {$j$} (X2.east);

          \draw (A.south) -- ++(-90:12pt) node[midway, right] {$n$};
          \draw (X1.west) -- ++(180:12pt) node[midway, above] {$\alpha$};
          \draw (X2.west) -- ++(180:12pt) node[midway, above] {$\beta$};
        \end{tikzpicture} \\
        M_{\alpha\beta\gamma}^{(n)} &= B_{ijk}^{(n)}X_{i\alpha}X_{j\beta}X_{k\gamma}
        =\begin{tikzpicture}[baseline={(0,0)}]
          \node[circle, minimum size=15pt, inner sep=0, fill=pink!60, draw=black] (B) at (1.25,0) {$B$};
          \node[circle, minimum size=15pt, inner sep=0, fill=orange!60, draw=black] (X1) at (0, 0.75) {$X$};
          \node[circle, minimum size=15pt, inner sep=0, fill=orange!60, draw=black] (X2) at (0, 0.0) {$X$};
          \node[circle, minimum size=15pt, inner sep=0, fill=orange!60, draw=black] (X3) at (0, -0.75) {$X$};

          \draw (B.north west) to node[midway, above] {$i$} (X1.east);
          \draw (B.west) to node[midway, above, yshift=-2pt] {$j$} (X2.east);
          \draw (B.south west) to node[midway, below] {$k$} (X3.east);

          \draw (B.south) -- ++(-90:12pt) node[midway, right] {$n$};
          \draw (X1.west) -- ++(180:12pt) node[midway, above] {$\alpha$};
          \draw (X2.west) -- ++(180:12pt) node[midway, above] {$\beta$};
          \draw (X3.west) -- ++(180:12pt) node[midway, above] {$\gamma$};
        \end{tikzpicture} \\
    \end{aligned}
\end{equation}

which, when substituted into Eq.~\eqref{eq:effective-hamiltonian}, fully defines the effective Hamiltonian as a function of the atomic configuration $\mathbf{X}$:

\begin{equation}\label{eq:effective-hamiltonian-full-penrose}
    \begin{aligned}
        \mathcal{H}_\text{eff}(\mathbf{X}) &= \frac{1}{2!}\varepsilon_{\alpha\beta}^{(n)}A_{ij}^{(n)}X_{i\alpha}X_{j\beta}\\
        &+ \frac{1}{3!}\zeta_{\alpha\beta\gamma}^{(n)}B_{ijk}^{(n)}X_{i\alpha}X_{j\beta}X_{k\gamma}\\
        &=\frac{1}{2!}\hspace{-7.5pt}\begin{tikzpicture}[baseline={(0,0)}]
            \node[circle, minimum size=15pt, inner sep=0, fill=red!30, draw=black] (E2) at (0, 0) {$\varepsilon$};
            \node[circle, minimum size=15pt, inner sep=0, fill=orange!60, draw=black] (X1) at (1.0, 0.5) {$X$};
            \node[circle, minimum size=15pt, inner sep=0, fill=orange!60, draw=black] (X2) at (1.0, -0.5) {$X$};
            \node[circle, minimum size=15pt, inner sep=0, fill=violet!60, draw=black] (A) at (2.0, 0.0) {$A$};

            \draw (E2.north east) to node[midway, above] {$\alpha$} (X1.west);
            \draw (E2.south east) to node[midway, below] {$\beta$} (X2.west);
            \draw (X1.east) to node[midway, above] {$i$} (A.north west);
            \draw (X2.east) to node[midway, below] {$j$} (A.south west);

            \draw (E2.south) .. controls +(-0.6,-1.0) and +(0.6,-1.0) .. node[midway,below]{$n$} (A.south);
        \end{tikzpicture} \\
        &+\frac{1}{3!}\hspace{-7.5pt}\begin{tikzpicture}[baseline={(0,0)}]
            \node[circle, minimum size=15pt, inner sep=0, fill=green!30, draw=black] (E3) at (0, 0) {$\zeta$};
            \node[circle, minimum size=15pt, inner sep=0, fill=orange!60, draw=black] (X1) at (1.0, 0.75) {$X$};
            \node[circle, minimum size=15pt, inner sep=0, fill=orange!60, draw=black] (X2) at (1.0, 0.0) {$X$};
            \node[circle, minimum size=15pt, inner sep=0, fill=orange!60, draw=black] (X3) at (1.0, -0.75) {$X$};
            \node[circle, minimum size=15pt, inner sep=0, fill=pink!60, draw=black] (B) at (2.0, 0.0) {$B$};

            \draw (E2.north east) to node[midway, above, yshift=4pt] {$\alpha$} (X1.west);
            \draw (E2.east) to node[midway, above, yshift=-2pt] {$\beta$} (X2.west);
            \draw (E2.south east) to node[midway, below, yshift=-2pt] {$\gamma$} (X3.west);
            \draw (X1.east) to node[midway, above, yshift=4pt] {$i$} (B.north west);
            \draw (X2.east) to node[midway, above, yshift=-2pt] {$j$} (B.west);
            \draw (X3.east) to node[midway, below, yshift=-2pt] {$k$} (B.south west);

            \draw (E3.south) .. controls +(-0.6,-1.25) and +(0.6,-1.25) .. node[midway,below]{$n$} (B.south);
        \end{tikzpicture}\hspace{-7.5pt}
    \end{aligned}
\end{equation}

The Penrose graphical notation used in Eq.~\eqref{eq:effective-hamiltonian-full-penrose} here clearly separates the effective Hamiltonian into three parts: learnable interaction parameters $\varepsilon$ and $\zeta$ on the left, a dynamic configuration tensor $\mathbf{X}$ in the middle, and static topology tensors $A$ and $B$, with the number of ``$X$" nodes representing the corresponding number of atoms in an interaction, e.g. two ``$X$" nodes representing two-body interactions. Additionally, diagrams for predicting tensorial properties would look nearly identical to the diagrams above, but with free edges representing tensor components. For example, for predicting internal stress $\sigma$, including only two-body terms for simplicity, the diagrams look nearly identical:

\begin{equation}
    \begin{aligned}
        \sigma_{pq}(\mathbf{X}) &= \frac{1}{2!} \Sigma_{\alpha\beta pq}^{(n)} A_{ij}^{(n)}X_{i\alpha}X_{j\beta} \\
        &= \frac{1}{2!}\hspace{1pt}\begin{tikzpicture}[baseline={(0,0)}]
            \node[circle, minimum size=15pt, inner sep=0, fill=cyan!60, draw=black] (E2) at (0, 0) {$\Sigma$};
            \node[circle, minimum size=15pt, inner sep=0, fill=orange!60, draw=black] (X1) at (1.0, 0.5) {$X$};
            \node[circle, minimum size=15pt, inner sep=0, fill=orange!60, draw=black] (X2) at (1.0, -0.5) {$X$};
            \node[circle, minimum size=15pt, inner sep=0, fill=violet!60, draw=black] (A) at (2.0, 0.0) {$A$};

            \draw (E2.north east) to node[midway, above] {$\alpha$} (X1.west);
            \draw (E2.south east) to node[midway, below] {$\beta$} (X2.west);
            \draw (X1.east) to node[midway, above] {$i$} (A.north west);
            \draw (X2.east) to node[midway, below] {$j$} (A.south west);

            \draw (E2.north west) -- ++ (145:12pt) node[midway, above] {$p$};
            \draw (E2.south west) -- ++(215:12pt) node[midway, below] {$q$};

            \draw (E2.south) .. controls +(-0.6,-1.0) and +(0.6,-1.0) .. node[midway,below]{$n$} (A.south);
        \end{tikzpicture}
    \end{aligned}
\end{equation}

where $p$ and $q$ denote directions and $\Sigma_{\alpha\beta pq}^{(n)}$ denotes the learnable tensor. For any learnable property, this model can be casted as a simple linear problem. In general, however, it is advantageous to cast an arbitrarily-shaped tensor to a vector property when fitting, or equivalently flattening the tensorial property. For example, flattening stress using Voigt notation\cite{jfnye} yields:

\begin{equation}
    \sigma = \begin{pmatrix}
        \sigma_{11} & \sigma_{22} & \sigma_{33} & \sigma_{23} & \sigma_{13} & \sigma_{12}
    \end{pmatrix}^\intercal
\end{equation}

Or, as a tensor contraction with a fixed mapping tensor $L_{\nu pq}$:

\begin{equation}
    \begin{aligned}
    \sigma_\nu(\mathbf{X}) &= L_{\nu pq} \sigma_{pq}(\mathbf{X})\\
    &= \frac{1}{2!}\hspace{1pt}\begin{tikzpicture}[baseline={(0,0)}]
            \node[circle, minimum size=15pt, inner sep=0, fill=cyan!60, draw=black] (E2) at (0, 0) {$\Sigma$};
            \node[circle, minimum size=15pt, inner sep=0, fill=orange!60, draw=black] (X1) at (1.0, 0.5) {$X$};
            \node[circle, minimum size=15pt, inner sep=0, fill=orange!60, draw=black] (X2) at (1.0, -0.5) {$X$};
            \node[circle, minimum size=15pt, inner sep=0, fill=violet!60, draw=black] (A) at (2.0, 0.0) {$A$};
            \node[circle, minimum size=15pt, inner sep=0, fill=yellow!90, draw=black] (L) at (-1.0, 0.0) {$L$};

            \draw (E2.north east) to node[midway, above] {$\alpha$} (X1.west);
            \draw (E2.south east) to node[midway, below] {$\beta$} (X2.west);
            \draw (X1.east) to node[midway, above] {$i$} (A.north west);
            \draw (X2.east) to node[midway, below] {$j$} (A.south west);

            \draw (E2.north west) to node[midway, above] {$p$} (L.north east);
            \draw (E2.south west) to node[midway, below] {$q$} (L.south east);

            \draw (L.west) -- ++ (180:12pt) node[midway, above] {$\nu$};

            \draw (E2.south) .. controls +(-0.6,-1.0) and +(0.6,-1.0) .. node[midway,below]{$n$} (A.south);
        \end{tikzpicture} \\
        &= \frac{1}{2!}\hspace{1pt}\begin{tikzpicture}[baseline={(0,0)}]
            \node[circle, minimum size=15pt, inner sep=0, fill=green!60, draw=black] (E2) at (0, 0) {$\Theta$};
            \node[circle, minimum size=15pt, inner sep=0, fill=orange!60, draw=black] (X1) at (1.0, 0.5) {$X$};
            \node[circle, minimum size=15pt, inner sep=0, fill=orange!60, draw=black] (X2) at (1.0, -0.5) {$X$};
            \node[circle, minimum size=15pt, inner sep=0, fill=violet!60, draw=black] (A) at (2.0, 0.0) {$A$};

            \draw (E2.north east) to node[midway, above] {$\alpha$} (X1.west);
            \draw (E2.south east) to node[midway, below] {$\beta$} (X2.west);
            \draw (X1.east) to node[midway, above] {$i$} (A.north west);
            \draw (X2.east) to node[midway, below] {$j$} (A.south west);

            \draw (E2.west) -- ++ (180:12pt) node[midway, above] {$\nu$};

            \draw (E2.south) .. controls +(-0.6,-1.0) and +(0.6,-1.0) .. node[midway,below]{$n$} (A.south);
        \end{tikzpicture}
    \end{aligned}
\end{equation}

where the new learnable tensor $\Theta$ is:

\begin{equation}
    \Theta_{\nu\alpha\beta}^{(n)} = L_{\nu pq}\Sigma_{\alpha\beta pq}^{(n)} = \hspace{1pt}\begin{tikzpicture}[baseline={(0,0)}]
            \node[circle, minimum size=15pt, inner sep=0, fill=cyan!60, draw=black] (E2) at (0, 0) {$\Sigma$};
            \node[circle, minimum size=15pt, inner sep=0, fill=yellow!90, draw=black] (L) at (-1.0, 0.0) {$L$};

            %\draw (E2.north east) to node[midway, above] {$\alpha$} (X1.west);
            %\draw (E2.south east) to node[midway, below] {$\beta$} (X2.west);

            \draw (E2.north west) to node[midway, above] {$p$} (L.north east);
            \draw (E2.south west) to node[midway, below] {$q$} (L.south east);

            \draw (L.west) -- ++ (180:12pt) node[midway, above] {$\nu$};
            \draw (E2.north east) -- ++ (30:12pt) node[midway, above] {$\alpha$};
            \draw (E2.south east) -- ++ (330:12pt) node[midway, below] {$\beta$};

            %\draw (E2.south) .. controls +(-0.6,-1.0) and +(0.6,-1.0) .. node[midway,below]{$n$} (A.south);
            \draw (E2.south) -- ++ (270:15pt) node[midway, below, yshift=-7.5pt] {$n$};
        \end{tikzpicture}
\end{equation}

Note that, in general, such a tensor $L$ needs to be invertible for the above transformation to be meaningful. Here, invertibility means that there is a one-to-one mapping between a tensor and its flattened version, which is true for the Voigt form above. We have implemented routines to train a CE model on a tensorial property in \verb|tce-lib|. However, for the sake of simplicity, we will focus on computing energy in this study.

We can decompose the model for the effective Hamiltonian (Eq.~\eqref{eq:effective-hamiltonian-full-penrose}) into a learnable vector parameter $\mathbf{j}$ and a feature vector $\mathbf{t}$:

\begin{equation}
    \mathcal{H}_\text{eff}(\mathbf{X}) = \mathbf{j}\cdot\mathbf{t}(\mathbf{X})
\end{equation}

where $\mathbf{t}(\mathbf{X})$ contains counts of cluster types:

\begin{equation}
    \mathbf{t}(\mathbf{X}) = \begin{pmatrix}
        \text{vec}(\mathbf{N}) \\ \text{vec}(\mathbf{M})
    \end{pmatrix}
\end{equation}

and $\mathbf{j}$ is a learnable interaction vector, analogous to the ECI from the traditional CE formalism:

\begin{equation}
    \mathbf{j} = \begin{pmatrix}
        \text{vec}(\boldsymbol{\varepsilon}) / 2! \\ \text{vec}(\boldsymbol{\zeta}) / 3!
    \end{pmatrix}
\end{equation}

and $\text{vec}$ denotes vectorization, or equivalently the flattening of a tensor into a column vector. Note that the computation of the $B$ elements depends on the structure of the lattice. For example, in a face-centered cubic (fcc) solid:

\begin{equation}
    B_{ijk}^{(1)} = \begin{cases}
        1 & A_{ij}^{(1)} = A_{jk}^{(1)} = A_{ki}^{(1)} = 1 \\
        0 & \text{else}
    \end{cases}
\end{equation}

or, equivalently, sites $i$, $j$, and $k$ form a first order triplet if all sites are connected via first nearest neighbor interactions. In the language of graphs, each triplet $(i, j, k)$ corresponds to a $3$-clique, or a complete induced subgraph of size $3$, of the graph defined by the adjacency tensor $A_{ij}^{(1)}$.

This can be generalized to a set of ``labels" per lattice structure. In the case of a first-order triplet in fcc, the corresponding label is $(1, 1, 1)$, denoting three first nearest-neighbor interaction. The second-order triplet in fcc, however, has a label of $(1, 1, 2)$, denoting that three sites form a second-order triplet if there are two first-nearest neighbors interactions, and one second-nearest neighbor interaction:

\begin{equation}
    B_{ijk}^{(2)} = \begin{cases}
        1 & A_{ij}^{(1)} = A_{jk}^{(1)} = A_{ki}^{(2)} = 1 \\
        1 & A_{ij}^{(1)} = A_{jk}^{(2)} = A_{ki}^{(1)} = 1 \\
        1 & A_{ij}^{(2)} = A_{jk}^{(1)} = A_{ki}^{(1)} = 1 \\
        0 & \text{else}
    \end{cases}
\end{equation}

where there is one conditional per unique permutation of $(1, 1, 2)$. More generally, for an order $n$ and a corresponding label $(\mathfrak{a}, \mathfrak{b}, \mathfrak{c})$:

\begin{equation}\label{eq:tensor-prod}
    B_{ijk}^{(n)} = \bigvee_{\sigma\in S_3} A_{ij}^{(\sigma(\mathfrak{a}))}A_{jk}^{(\sigma(\mathfrak{b}))}A_{ki}^{(\sigma(\mathfrak{c}))}
\end{equation}

where $S_3$ denotes the symmetric group on $3$ elements and $\vee: \{0, 1\}^2 \to \{0, 1\}$ denotes the logical OR operation, i.e. we are performing a Boolean sum over the permutations of $(\mathfrak{a}, \mathfrak{b}, \mathfrak{c})$, and $i$, $j$, and $k$ are free indices. The mapping between orders and labels is entirely a function of the crystal structure of the system at hand; a more comprehensive table of these labels is provided in Table \ref{tab:three-body-labels}.

\begin{table}[H]
    \centering
    \begin{tabular}{|c|c|c|c|}
\hline
\diagbox{label}{structure} & sc & bcc & fcc \\ \hline
$(1, 1, 1)$ &  &  & \checkmark \\ \hline
$(1, 1, 2)$ & \checkmark & \checkmark & \checkmark \\ \hline
$(1, 1, 3)$ &  & \checkmark & \checkmark \\ \hline
$(1, 2, 2)$ &  &  &  \\ \hline
$(1, 2, 3)$ & \checkmark &  & \checkmark \\ \hline
$(1, 3, 3)$ &  &  & \checkmark \\ \hline
$(2, 2, 2)$ & \checkmark &  &  \\ \hline
$(2, 2, 3)$ &  & \checkmark &  \\ \hline
$(2, 3, 3)$ &  &  & \checkmark \\ \hline
$(3, 3, 3)$ &  & \checkmark & \checkmark \\ \hline
\end{tabular}

\iffalse
from tce import constants

def generate_latex_table(S: set[tuple[int, int, int]], mapping: dict[constants.LatticeStructure, list[tuple[int, int, int]]]) -> str:
    # Sort the elements for consistency
    elements = sorted(S)
    headers = [letter.name.lower() for letter in constants.LatticeStructure]

    lines = [
        r"\begin{tabular}{|c" + "|c" * len(constants.LatticeStructure) + "|}",
        r"\hline",
        r"\diagbox{label}{structure} & " + " & ".join(headers) + r" \\ \hline"
    ]

    for elem in elements:
        row = [f"${elem}$"]
        for letter in constants.LatticeStructure:
            if elem in mapping.get(letter, []):
                row.append(r"\checkmark")
            else:
                row.append("")
        lines.append(" & ".join(row) + r" \\ \hline")
    
    # End table and document
    lines += [
        r"\end{tabular}"
    ]

    return "\n".join(lines)

def main():

    max_label_int = 2

    possible_labels = set()
    for i in range(max_label_int + 1):
        for j in range(i + 1):
            for k in range(j + 1):
                possible_labels.add((i + 1, j + 1, k + 1))

    possible_labels = [tuple(sorted(x)) for x in possible_labels]
    possible_labels.sort(key=lambda x: (max(x), x))

    labels = {
        structure: set(
            tuple(int(k) + 1 for k in x) for x in constants.STRUCTURE_TO_THREE_BODY_LABELS[structure]
        ) for structure in constants.LatticeStructure
    }
    
    print(generate_latex_table(possible_labels, labels))

if __name__ == "__main__":

    main()

\fi

    \caption{Three-body labels for triplets for sc, bcc, and fcc crystal structures, up to third nearest neighbors.}
    \label{tab:three-body-labels}
\end{table}

Note that calculating a new mapping between orders and labels for a new lattice type for three body terms is trivial, i.e. pick a tuple of labels, and calculate the three-body topology tensor for that tuple of labels using Eq.~\eqref{eq:tensor-prod}. If this tensor product is non-zero, then this tuple of labels is a valid three-body term to be included in $\mathcal{H}_\text{eff}$ for that lattice type. Additionally, this computation needs to only be performed once on a small lattice, and can then be stored in a database such as Table \ref{tab:three-body-labels} to be later used on a larger lattice. Although Table \ref{tab:three-body-labels} only includes sc, bcc, and fcc lattices, this process generalizes to any periodic lattice structure. This routine has been implemented in \verb|tce-lib| for any cubic lattice structure with an arbitrary atomic basis. This routine will additionally be generalized to non-cubic lattice structures in a future work.

Additionally, higher-order interactions would be similarly encoded by a tensor product like Eq.~\ref{eq:tensor-prod}, with a set of labels depending on structure. However, in this work, we only use two and three body interactions, and therefore the enumeration of higher-order labels is saved for a future work.

Furthermore, note that $\mathcal{H}_\text{eff}$ is linear in $\mathbf{t}$. Therefore, we can efficiently learn an interaction vector $\mathbf{j}$ from data-pairs $(\mathbf{t}, \mathcal{H}_\text{eff})$ using one of the various linear regression techniques, from atomistic data. However, care must be taken in the fitting routine since the feature vector $\mathbf{t}$ has redundant features. For example, $\varepsilon_{12}^{(1)} = \varepsilon_{21}^{(1)}$, but both $N_{12}^{(1)}$ and $N_{21}^{(1)}$ are included as features. This is traditionally remedied via regularization - typically lasso, ridge, or elastic net. In this work, we choose a regularization routine that is equivalent to ridge regularization, but for an arbitrarily small regularization parameter, which is described further in Section \ref{sec:training}.

An additional advantage of this formalism is that the topology of the lattice is explicitly embedded within the effective Hamiltonian. This means that the local energies are a direct consequence of the formalism, which reduces the necessary computational resources to compute energy differences. The difference in energy between two states $\mathbf{X}$ and $\mathbf{X}'$ can be written as:

\begin{equation}
    \mathcal{H}_\text{eff}(\mathbf{X}') - \mathcal{H}_\text{eff}(\mathbf{X}) = \mathbf{j}\cdot \left(\mathbf{t}(\mathbf{X}') - \mathbf{t}(\mathbf{X})\right)
\end{equation}

Note that, for two ``close" states $\mathbf{X}$ and $\mathbf{X}'$, most terms in the contractions to compute $\mathbf{t}(\mathbf{X}') - \mathbf{t}(\mathbf{X})$ are redundant. Physically, this is because two ``close" states can be formed via a small number of atomic swaps, only changing a small number of local energies. This is especially important, since energy differences are the quantity of interest in an atomistic simulation, and recomputing redundant energies far away from the atomic swaps is wasteful.

Since the feature vector $\mathbf{t}$ has explicitly-encoded topology (in the form of spatial indices $i$, $j$, $\cdots$), this can be accounted for analytically. Let $\mathcal{D}$ be the set of lattice sites with differing atom types:

\begin{equation}
    \mathcal{D} = \{i: \mathbf{X}_i \neq \mathbf{X}_i'\}
\end{equation}

Then, we simply only include terms in the contraction where any of the spatial indices appears in $\mathcal{D}$. Equivalently, we can truncate the contraction along a single spatial index, and symmetrize:

\begin{equation}\label{eq:shortcut}
    \begin{aligned}
        \Delta \tilde{N}_{\alpha\beta}^{(n)} &= \sum_{d\in\mathcal{D}}A_{dj}^{(n)}X'_{d\alpha}X'_{j\beta}\\
        &- \sum_{d\in\mathcal{D}}A_{dj}^{(n)}X_{d\alpha}X_{j\beta} \\
        \Delta N_{\alpha\beta}^{(n)} &= 2\Delta\tilde{N}_{(\alpha\beta)}^{(n)} \\
        \Delta\tilde{M}_{\alpha\beta\gamma}^{(n)} &= \sum_{d\in\mathcal{D}} B_{djk}^{(n)}X_{d\alpha}'X_{j\beta}'X_{k\gamma}'\\
        &- \sum_{d\in\mathcal{D}} B_{djk}^{(n)}X_{d\alpha}X_{j\beta}X_{k\gamma} \\
        \Delta M_{\alpha\beta\gamma}^{(n)} &= 3\Delta \tilde{M}_{(\alpha\beta\gamma)}^{(n)}
    \end{aligned}
\end{equation}

yielding:

\begin{equation}
    \Delta\mathbf{t} = \mathbf{t}(\mathbf{X}') - \mathbf{t}(\mathbf{X}) = \begin{pmatrix}\text{vec}(\Delta\mathbf{N}) \\ \text{vec}(\Delta\mathbf{M})\end{pmatrix}
\end{equation}

where parenthesis around lower indices denotes symmetrization of an arbitrary rank-$r$ tensor $\boldsymbol{\Xi}$:

\begin{equation}
    \Xi_{(i_1 i_2 \cdots i_r)} = \frac{1}{r!}\sum_{\sigma\in S_r} \Xi_{\sigma(i_1) \sigma(i_2) \cdots \sigma(i_r)}
\end{equation}

and $S_r$ denotes the symmetric group on $r$ elements. In the naive approach, the number of terms needed to compute $\Delta\mathbf{t}$ is $\mathcal{O}(N^3)$, where $N$ is the number of lattice sites in the supercell. With this shortcut, the number of terms is reduced instead to $\mathcal{O}(|\mathcal{D}|N^2)$. If $A_{ij}^{(n)}$ and $B_{ijk}^{(n)}$ are stored sparsely, these can further be reduced to $\mathcal{O}(N)$ and $\mathcal{O}(|\mathcal{D}|)$ for the naive and shortcut methods, respectively.

Various popular CE libraries, including but not necessarily limited to ICET\cite{icet} and CASM \cite{casm}, already include similar local, incremental energy differences within their Monte Carlo (MC) routines. However, the correlation function differences must be treated on a case-by-case basis. Here, since the incremental update is a natural consequence of our TCE formalism, the energy update implementation is universal for any lattice structure, any set of chemical species, and any set of cluster types, including clusters on exotic lattices. Secondly, a tensor contraction is more friendly for massively parallelized architectures, such as GPUs or TPUs. Lastly, $\mathcal{H}_\text{eff}$ is explicitly written as a multilinear form in binary values, and can therefore be well-optimized using classical integer programming\cite{Balas1984} or quantum annealers\cite{10820320}.

\subsection{Training}\label{sec:training}

Suppose we have a configuration space $\mathcal{C} = \{0, 1\}^{N\times m}$, where $N$ is the number of lattice sites and $m$ is the number of alloying types, and an oracle, i.e. a black-box function mapping configuration to energy, $f: \mathcal{C}\to \mathbb{R}$. $f$ is typically a DFT calculation of the configuration's energy, but could be any energy-computing method in principle.

Then, we sample configuration space by generating a sequence of $I$ configurations $(\mathbf{X}_1, \cdots, \mathbf{X}_I)$, and compute their energies $(f(\mathbf{X}_1), \cdots, f(\mathbf{X}_I)) = (E_1, \cdots, E_I)$.

Then, we compute the corresponding feature vectors for each sample $s\in [1, I]$:

\begin{equation}
    \begin{aligned}
        \mathbf{X} &= \mathbf{X}_s \\
        N_{\alpha\beta}^{(n)} &= A_{ij}^{(n)}X_{i\alpha}X_{j\beta} \\
        M_{\alpha\beta\gamma}^{(n)} &= B_{ijk}^{(n)}X_{i\alpha}X_{j\beta}X_{k\gamma} \\
        \mathbf{t}_s &= \begin{pmatrix}
            \text{vec}(\mathbf{N}) \\ \text{vec}(\mathbf{M})
        \end{pmatrix}
    \end{aligned}
\end{equation}

and define a corresponding ridge regression loss with regularization parameter $\lambda > 0$:

\begin{equation}
    \begin{aligned}
        \mathbf{T} &= \begin{pmatrix}
            \mathbf{t}_1 & \cdots & \mathbf{t}_I
        \end{pmatrix} \\
        \mathbf{e} &= \begin{pmatrix}
            E_1 & \cdots & E_I
        \end{pmatrix}^\intercal \\
        L(\mathbf{j}\;|\;\lambda) &= \left\|\mathbf{T}\mathbf{j} - \mathbf{e}\right\|^2 + \lambda\left\|\mathbf{j}\right\|^2
    \end{aligned}
\end{equation}

and then minimize this loss in the limit that the regularization term goes to $0$:

\begin{equation}
    \begin{aligned}
        \mathbf{j}_\text{opt}(\lambda) &= \arg\min_{\mathbf{j}}L(\mathbf{j}\;|\;\lambda)\\
        &= \left(\mathbf{T}^\intercal\mathbf{T} + \lambda\mathbf{I}\right)^{-1}\mathbf{T}^\intercal\mathbf{e}\\
        \mathbf{j}^* &= \lim_{\lambda\to 0^+}\mathbf{j}_\text{opt}(\lambda) = \mathbf{T}^+ \mathbf{e}
    \end{aligned}
\end{equation}

where $\mathbf{T}^+$ is the Moore-Penrose inverse, which can be computed using singular value decomposition of $\mathbf{T}$\cite{moore-penrose-limit}.

Note that, if $\mathbf{T}$ is full-rank, i.e. all features are linearly independent, then $\mathbf{T}^+ = \left(\mathbf{T}^\intercal\mathbf{T}\right)^{-1}\mathbf{T}^\intercal$. However, this is not true in the optimization problem above. Since the tensors $\mathbf{N}$ and $\mathbf{M}$ are symmetric, $\mathbf{T}$ is not full-rank, yielding a singular $\mathbf{T}^\intercal\mathbf{T}$. Despite this, however, the Moore-Penrose inverse is still well-defined. Heuristically, this is because the singular values of zero correspond to redundant data in $\mathbf{T}$ (the rank of $\mathbf{T}$ is exactly equal to the number of non-zero singular values), and these singular values of zero are ignored in the Moore-Penrose inversion.

Additionally, this fitting routine ensures necessary symmetries in the fitting parameters $\varepsilon_{\alpha\beta}^{(n)}$ and $\zeta_{\alpha\beta\gamma}^{(n)}$, namely that $\varepsilon_{\alpha\beta}^{(n)} = \varepsilon_{(\alpha\beta)}^{(n)}$ and $\zeta_{\alpha\beta\gamma}^{(n)} = \zeta_{(\alpha\beta\gamma)}^{(n)}$. More generally, if $\mathbf{T}$ is the data matrix, $\mathbf{j} = \mathbf{T}^+ \mathbf{e}$ is the fitted model parameter, and the $a$'th and $b$'th columns of $\mathbf{T}$ are equal, then $j_a = j_b$ \cite{moore-penrose-feature-equivalence}.
\section{Methods and Results}

To test the methodology above, we perform three numerical experiments: an empirical analysis of the computation saved via the shortcut to compute $\Delta\mathbf{t}$, calculation of the heat of mixing curve for the bcc $\text{Ta}_{1-x}\text{W}_x$ binary alloy from DFT data using Vienna Ab Initio Simulation Package (VASP)\cite{vasp1,vasp2}, and calculation of the equilibrium Cowley short range order (SRO) parameters\cite{cowley-sro-paper} of the equiatomic fcc CoNiCrFeMn alloy from a MEAM potential created by Lee et. al\cite{cantor-potential}.

\subsection{Timing benchmark}

The first numerical experiment is to benchmark the computational resources saved by using our aforementioned shortcut to compute $\Delta\mathbf{t}$ vs. computing it naively. This is done by building a fcc supercell with a sequence of various sizes. For each size, we initialize a random equiatomic ABC configuration $\mathbf{X}$, and randomly swap two atoms to create a different configuration $\mathbf{X}'$. Then, we compute $\Delta\mathbf{t}$ naively, and with the aforementioned shortcut, and record the timings for both computations. We repeat this for multiple pairs of two-body and three-body orders, i.e. how many two-body and three-body terms are included in the computation.

\begin{figure}[H]
    \centering
    \includegraphics[width=\linewidth]{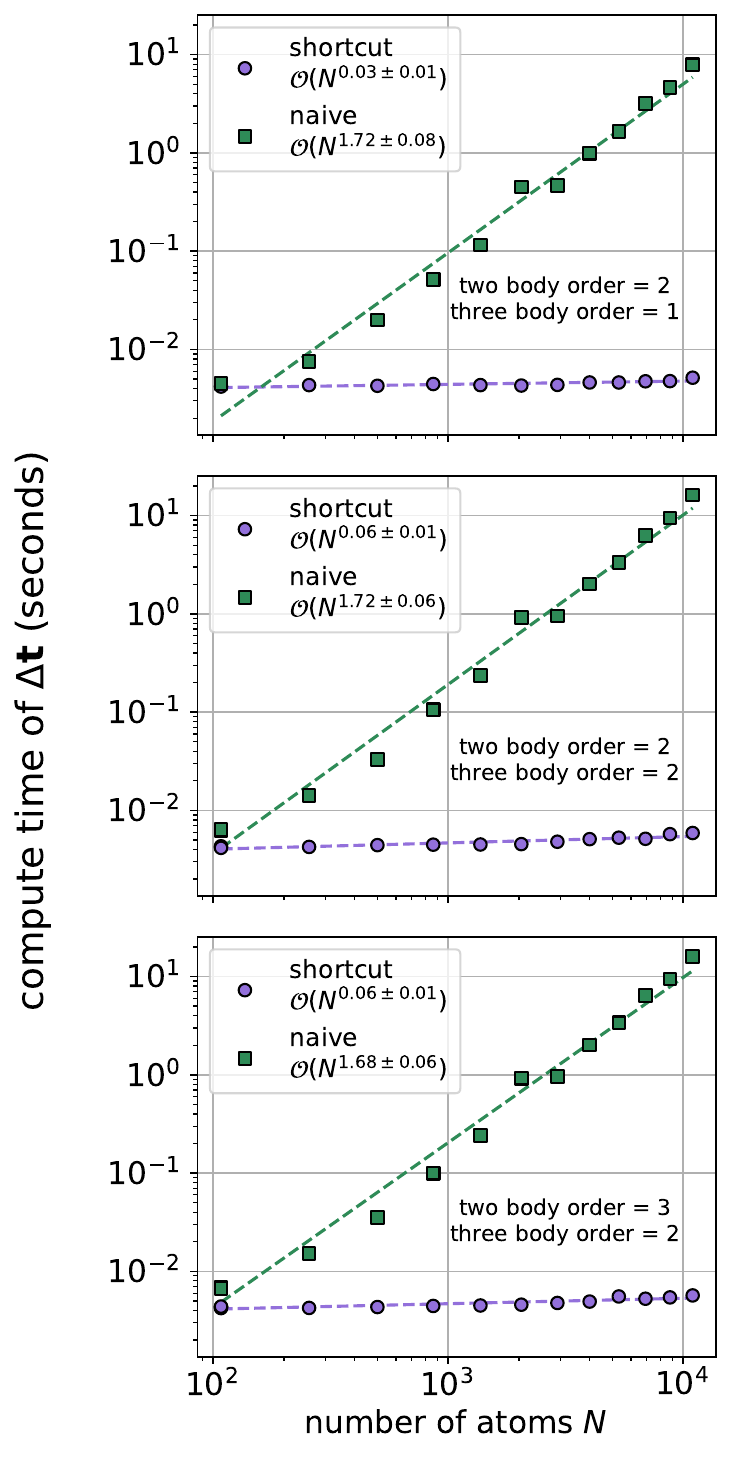}
    \caption{Naive vs. shortcut computation time benchmarked over various system sizes with respective power law fits. The time complexities presented in the legends are determined via power-law fits (denoted by dashed lines) to the presented data points.}
    \label{fig:shortcut}
\end{figure}

Below, we showcase speedups of computational time when computing $\Delta\mathbf{t}$ with our aforementioned shortcut derived in Eq.~\eqref{eq:shortcut} and without. When computing the feature difference between two states naively, i.e. computing the full feature vector for both states, the computation time scales roughly with $N^{1.7}$, where $N$ denotes the system size. However, when computing the feature difference with our shortcut, the computation time scales roughly with $N^{0.05}$ (Figure \ref{fig:shortcut}). Notably, this is not exactly $\mathcal{O}(1)$ as theoretically predicted. We expect that this is due to the time necessary to generate a copy of a sub-array of a \verb|sparse| Coordinate List (COO) array, which likely grows with array size.

The empirical time complexity additionally seems to be largely independent of the expansion order, for both the naive and shortcut computations of $\Delta t$. Counter-intuitively, the best-fit exponent for the naive method seems to slightly decrease with expansion order. However, we note that this decrease is smaller than the corresponding uncertainty of the fits, and we expect that this is a statistical artifact.

Here, our shortcut computation time is largely independent of system size, indicating that the computation time of an MC step is constant for an arbitrarily large system. We do note that, with tce-lib, this trend seems to break for systems larger than about 20,000 atoms. We expect that this is entirely implementation-dependent, and plan to further optimize \verb|tce-lib| for larger systems. The empirical scaling relationship shown here is nevertheless very promising for simulating larger atomistic systems. Furthermore, the absolute computation times are still quite large (larger than $\SI{1}{ms}$). However, this is implementation and hardware-dependent, and we hope to further optimize this in a future work.

Lastly, we note that these time complexities are calculated purely for the computation of feature differences $\Delta \mathbf{t}$, and that no CE models were trained in the numerical experiments above. These time complexities are relevant in the model deployment phase, e.g. running a MC simulation. In the training phase, however, the time necessary to train should increase with the expansion order.

\subsection{Heat of mixing of TaW}

The second numerical experiment is to showcase a prediction of an enthalpy of mixing curve in the bcc TaW binary system, which is of interest in fusion reactor applications due to the effect of Ta-alloying on W's mechanical and thermal properties\cite{NOGAMI2022153740,WURSTER2011166}. This is done by first generating a diverse training set via a genetic algorithm. We first start with five seed configurations (pure $\text{Ta}$, pure $\text{W}$, $\text{B}_2$ $\text{Ta}\text{W}$, $\text{DO}_3$ $\text{Ta}\text{W}_3$, and $\text{DO}_3$ $\text{Ta}_3\text{W}$), each with 128 atoms.

From these seeds, we randomly pick two configurations, one labeled a mother and one labeled a father. A child configuration is then created by taking half of the configuration from the mother configuration, and the other half from the father configuration, split evenly along either the $(100)$, $(010)$, or $(001)$ plane, which is also randomly selected. We then introduce random mutations, i.e. each lattice site in the child has a $1\%$ chance of being independently sampled. We note that new samples are generated purely via random crossover and mutation, and that manual selection is only necessary for choosing ordered parent structures.

This child is then added to the set of choose-able configurations, and this process is repeated $245$ times to create a total of $250$ training samples. For each sample, we compute the total energy with DFT using VASP, which is often used to train alloy CE models \cite{el-atwani_outstanding_2019}. VASP calculations were performed using the Atomic Simulation Environment (ASE)\cite{ase} using the projector augmented-wave method with Perdew-Burke-Ernzerhof exchange correlation functionals\cite{pbe} with Gaussian smearing with a width of $\SI{0.1}{\electronvolt}$ and energy cutoff $\SI{500}{eV}$ on a $\Gamma$-centered $3\times 3\times 3$ $k$-point mesh. Convergence thresholds for total energy and force were, respectively, $\SI{1e-5}{\electronvolt}$ and $\SI{5e-4}{\electronvolt\per\angstrom}$.

These energies are then used to train the CE model using the training routine described above using \verb|tce-lib|, including up to third nearest neighbors and the three-body terms with labels $(1, 1, 2)$ and $(1, 1, 3)$. We then compute the resulting enthalpy of mixing curve at $\SI{0}{K}$ by computing the interaction energy-per-atom from the trained CE model for $32$ different compositions on a $10\times 10\times 10$ bcc lattice:

\begin{equation}
    \Delta H_\text{mix}(x) = E(x) - xE_\text{W} - (1 - x)E_\text{Ta}
\end{equation}

where $x$ is the atomic fraction of $\text{W}$, and $E_\text{W}$ and $E_\text{Ta}$ are respectively the cohesive energy of pure $\text{W}$ and $\text{Ta}$, also computed with DFT. The interaction energy $E(x)$ is computed by initializing a random solid solution, and then equilibrating for $10^5$ steps using Metropolis MC\cite{metropolis-mc} at $\SI{300}{K}$ per composition $x$. At each MC step, we randomly swap two atoms, and accepting the swap if $\exp(-\beta\Delta E) > u$, where $u$ is a uniformly-sampled random number between $0$ and $1$. $\beta$ here denotes inverse temperature (rather than an alloying type), i.e. $\beta = 1/k_BT$.

To benchmark the accuracy of our enthalpy of mixing curve, we then compute the corresponding enthalpy of mixing obtained from a previously developed CE \cite{alvarado2023} run at 300 K
from a hybrid MC routine as implemented in ATAT \cite{atat}. The configurations obtained following this methodology are then fed into VASP and minimize at constant pressure to obtain the final energy, used to calculate the enthalpy of mixing.

Here, we present the results of both the training data diversity from the aforementioned genetic algorithm and the resulting learned interaction parameters for the TaW binary alloy, learned from DFT. We use the Cowley short-range order (SRO) parameter as a fingerprint for chemical ordering:

\begin{equation}\label{eq:cowley-sro}
    \chi_{\alpha\beta} = 1 - \frac{p_{\alpha\beta}}{(2 - \delta_{\alpha\beta})x_\alpha x_\beta}
\end{equation}

where $p_{\alpha\beta}$ is the probability of finding two chemical species $\alpha$ and $\beta$ in the first-nearest neighbor shell, $\delta_{\alpha\beta}$ is the Kronecker delta, and $x_\alpha$ is the atomic fraction of species $\alpha$ within the sample\cite{cowley-sro-paper}. The distribution of $\text{Ta}$-$\text{W}$ SRO parameters $\chi_{\text{Ta}\text{W}}$ has a large spread, which means that the resulting CE model is exposed to a wide diversity of chemical orderings during training (Figure \ref{fig:sro-dist-genetic}). We have additionally included a visualization of the entire training set in the Supplementary Materials.

\begin{figure}[H]
    \centering
    \includegraphics[width=0.6\linewidth]{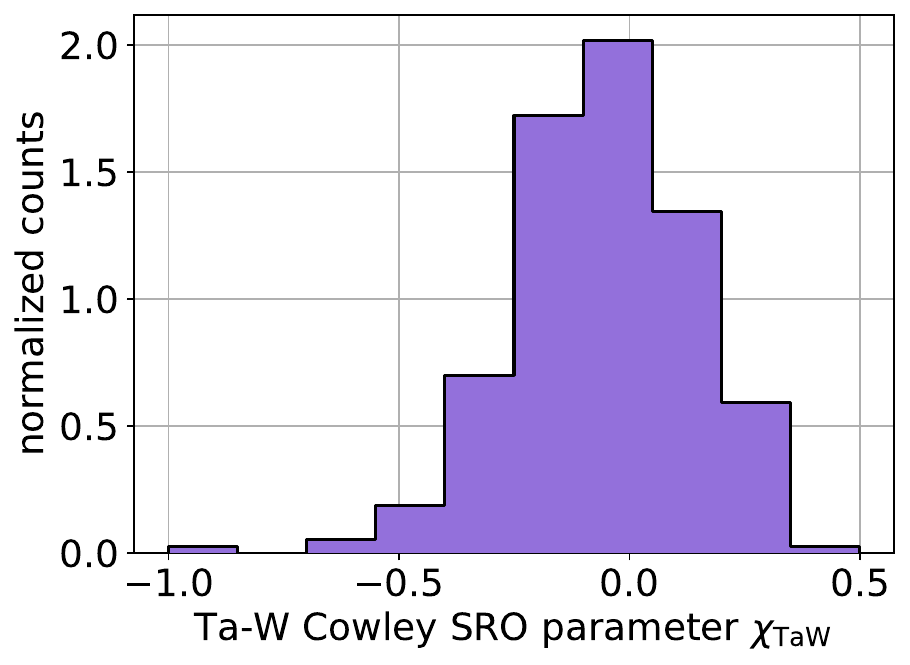}
    \includegraphics[width=0.35\linewidth]{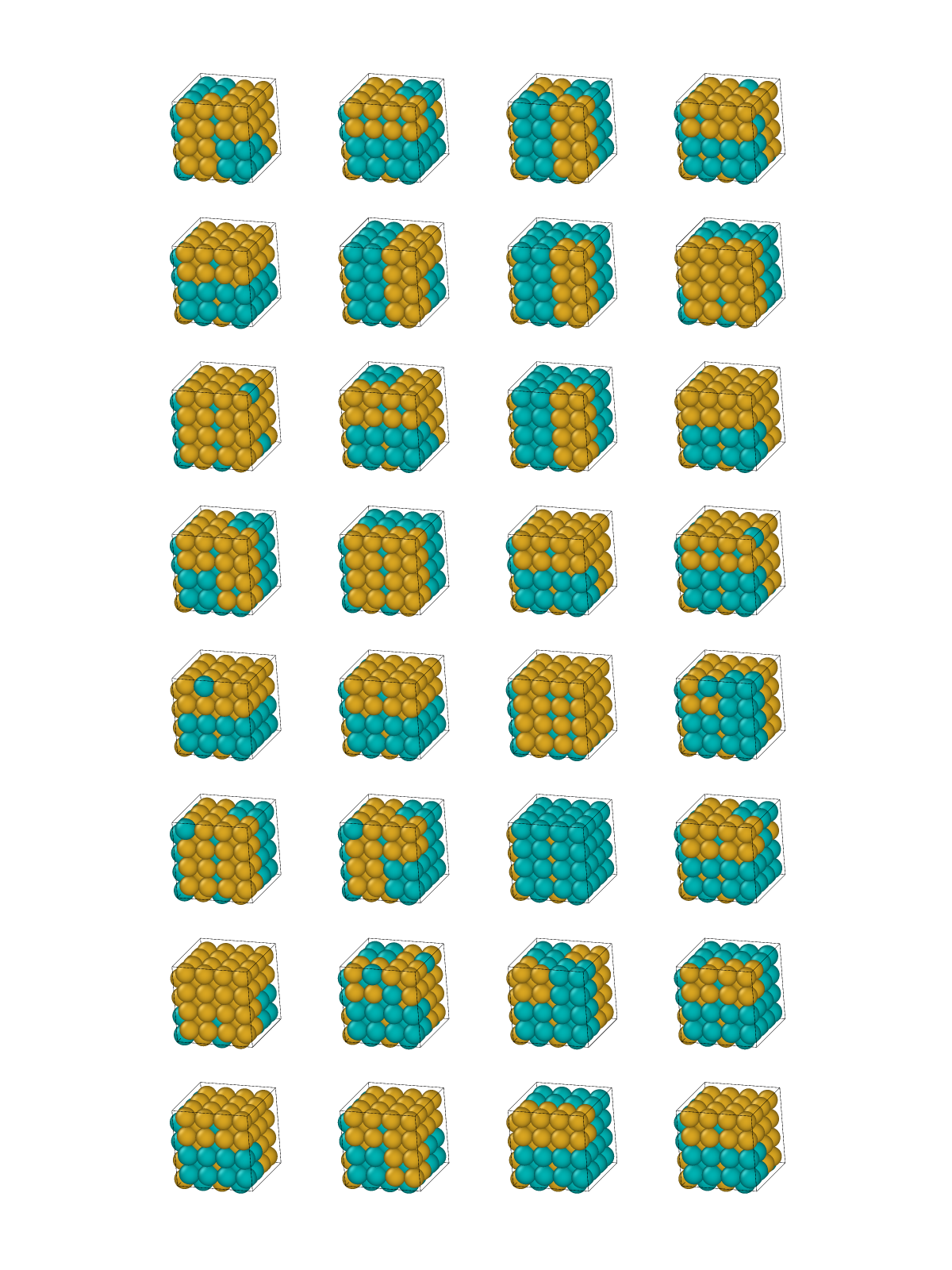}
    \rule{0.4\textwidth}{0.4pt}
    \includegraphics[width=\linewidth]{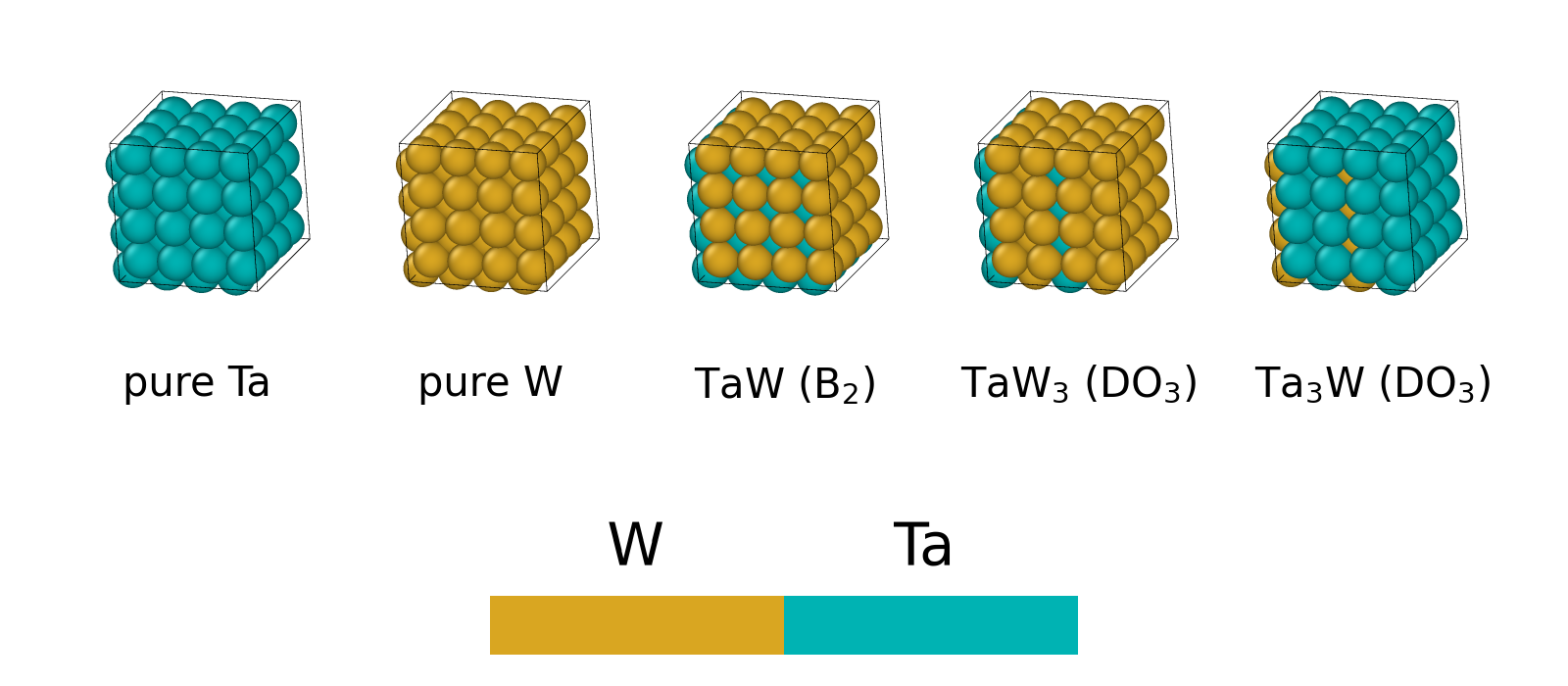}
    \caption{Distribution of SROs in training data generated from genetic algorithm (top left), and 32 example configurations (top right) out of the 500 total, starting from a set of ordered seeds (bottom). All atomic configurations are rendered using OVITO.}
    \label{fig:sro-dist-genetic}
\end{figure}

Here, $\chi_{\text{Ta}\text{W}} = -1$ denotes a B2 sample; i.e. every first-nearest neighbor of a $\text{Ta}$ atom is a $\text{W}$ atom, and vice versa, while positive SRO parameters denote phase separation. For a binary system, the generation of these extreme chemical orderings is relatively trivial. The genetic algorithm efficiently samples configurations between these extreme orderings, generating closer-to-random samples, with $\chi_{\text{Ta}\text{W}} \approx 0$. Essentially, the genetic algorithm starts from the ``corners" of chemical ordering space, smoothly bridging the gaps in between, efficiently sampling chemical ordering space.

\begin{figure}[H]
    \centering
    \includegraphics[width=\linewidth]{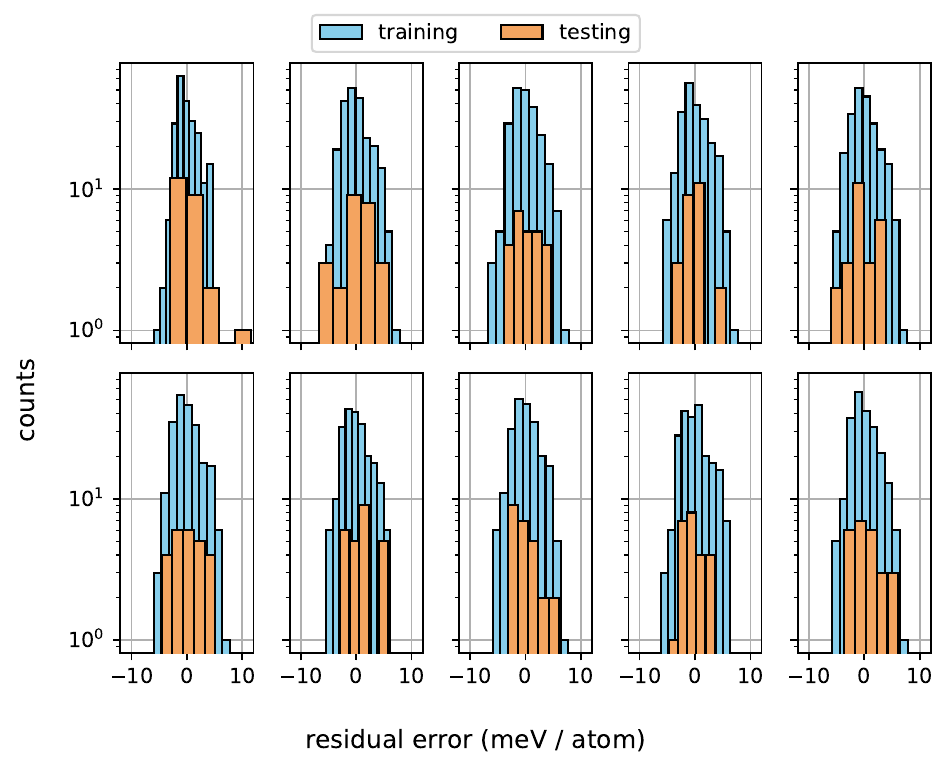}
    \caption{TaW cross validation residual errors. Each plot represents a different $90\%/10\%$ train-test split. One outlier (approximately $\SI{-40}{meV/atom}$) is excluded in the top left split.}
    \label{fig:taw-cross-val}
\end{figure}

\begin{figure}[H]
    \centering
    \includegraphics[width=\linewidth]{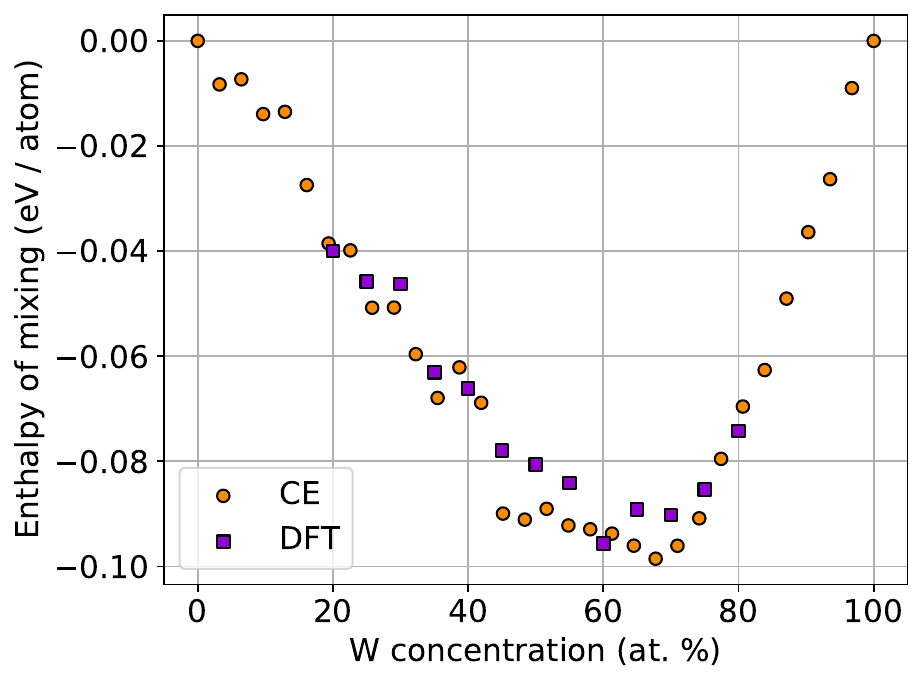}
    \caption{TaW enthalpy of mixing calculated via our CE and the DFT-MC routine.}
    \label{fig:taw-convex}
\end{figure}

This algorithm is not only very efficient, but also seems to expose the CE fitting routine to a training set that is sufficiently diverse to accurately predict both absolute energies (Figure \ref{fig:taw-cross-val}) and the enthalpy of mixing (Figure \ref{fig:taw-convex}) over a large composition range. When validated against DFT data using $10$-fold cross validation, the prediction errors are quite low in magnitude, on the order of $\SI{5}{meV/atom}$ to $\SI{10}{meV/atom}$ for our generated TaW data.

In each train-test split, both training and testing residual errors are centered at around $\SI{0}{meV/atom}$, with a spread of roughly $\SI{5}{meV/atom}$ to $\SI{10}{meV/atom}$. Since both training and testing residual errors are similar for each train/test split (in the distribution sense), we can be reasonably confident that the CE model does not overfit the DFT data. Additionally, since both errors are centered at zero and roughly symmetric about this center, we can also be confident that we have enough features (i.e. clusters) to well-capture the complexity of the training set.

Additionally, the CE model accurately predicts the enthalpy of mixing across the whole composition range. Both indicate a relatively strong tendency to mix, with a minimum enthalpy of mixing of $\SI{-0.08}{eV/atom}$ to $\SI{-0.10}{eV/atom}$ at around $60\%$-$70\%$ $\text{W}$ (Figure \ref{fig:taw-convex}). This is consistent with the experimental enthalpy of mixing curve calculated by Singhal and Worrell, which report a minimum enthalpy of mixing at approximately $40\%$ Ta \cite{Singhal1973}.

\subsection{Short range ordering of CoNiCrFeMn}

The final numerical experiment is to showcase the prediction of the full set of Cowley SRO parameters for the equiatomic fcc CoNiCrFeMn alloy computed via an Open Visualization Tool (OVITO)\cite{ovito} modifier\cite{jeffries2025prediction, jeffries_sro_parameters}. This is done by first generating a diverse training set via surrogate models. We first generate $500$ random pairwise, first nearest neighbor surrogate cluster expansion (CE) models. Interaction parameters are drawn from a normal distribution and then symmetrized, i.e. $\tilde{\varepsilon}_{\alpha\beta}^{(1)}\sim\mathcal{N}(\mu, \sigma^2)$ and $\varepsilon_{\alpha\beta}^{(1)} = \tilde{\varepsilon}_{(\alpha\beta)}^{(1)}$, where $\mu = \SI{-1.0}{\electronvolt}$ and $\sigma = \SI{0.3}{\electronvolt}$. These surrogate models are then used to relax small $4\times 4\times 4$ fcc systems in parallel, using canonical Metropolis Monte Carlo (MC).

In each MC step, we randomly swap two atoms of differing alloying types, compute the energy difference $\Delta E$, and then accept or reject the swap if $\exp(-\beta\Delta E) > u$, where $u$ is a uniformly-sampled random number between $0$ and $1$. $\beta$ is additionally randomly sampled for each surrogate system between $\SI{5.0}{\per\electronvolt}$ and $\SI{40.0}{\per\electronvolt}$, corresponding to approximate temperatures of $\SI{2320}{K}$ and $\SI{290}{K}$. We note that these surrogate systems are completely unphysical, and are only a means to generate a configurationally diverse training set for a CE model.

We then compute each surrogate system's ``true" energy with a MEAM potential using Large-scale Atomic/Molecular Massively Parallel Simulator (LAMMPS)\cite{LAMMPS}. We note that the resulting CE model is likely not as quantitatively accurate, since interatomic potentials are generally less accurate than DFT. However, we choose here to use an interatomic potential to showcase that our routine not only works for DFT, but with other energy-calculating drivers as well.

These energies and configurations are then similarly used to train a CE model using \verb|tce-lib|, including up to second nearest-neighbors and one three body term with label $(1, 1, 1)$. We then use this trained model to equilibrate a $10\times 10\times 10$ fcc CoNiCrFeMn lattice using Metropolis MC, and track the Cowley SRO parameters throughout the run.

We then perform a similar MC/MD routine using LAMMPS to benchmark the accuracy of the CE-predicted SRO parameters. In this routine, we run $10^6$ MD steps with a $\SI{1}{fs}$ timestep with a Nose-Hoover thermostat and barostat set to $\SI{600}{K}$ and $\SI{1.0}{bar}$ with respective relaxation times $\SI{0.1}{ps}$ and $\SI{1.0}{ps}$. Every $10^3$ MD steps, we attempt $5$ swaps per type pair, and accept or reject according to the Metropolis criterion.

Similar results are shown here on the diversity of the training set generated via first-nearest neighbor pairwise surrogate CE models and the resulting interaction parameters learned from a MEAM potential. The distribution of SRO parameters has again a large spread for all pairs of alloying types, indicative of ordering diversity within the training set (Figure \ref{fig:sro-dist-surrogate}). We have similarly included a visualization of the entire training set in the Supplementary Materials.

\begin{figure}[H]
    \centering
    \includegraphics[width=\linewidth]{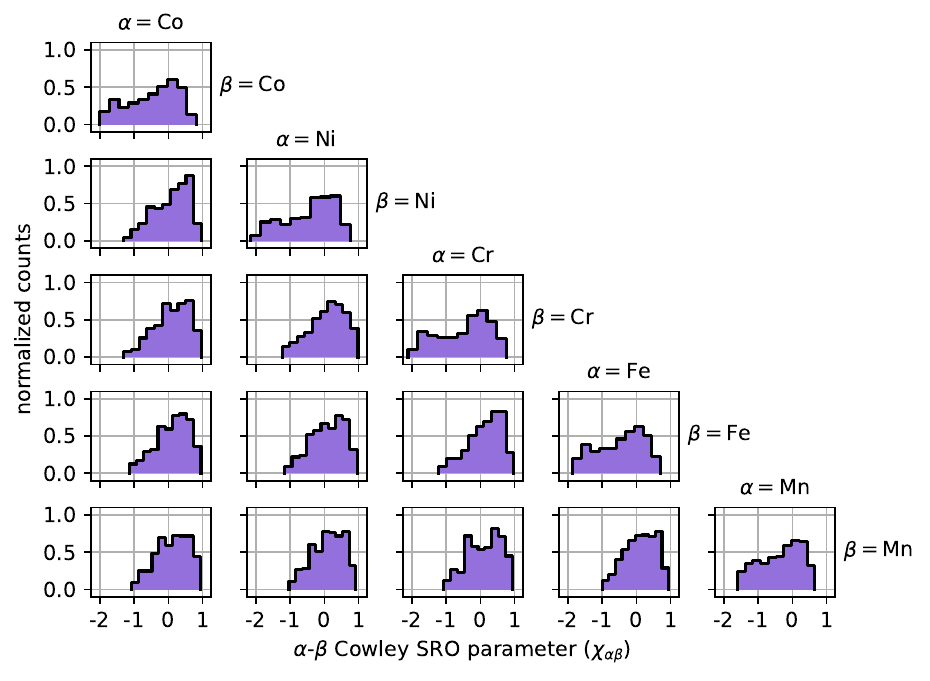}
    \includegraphics[width=\linewidth]{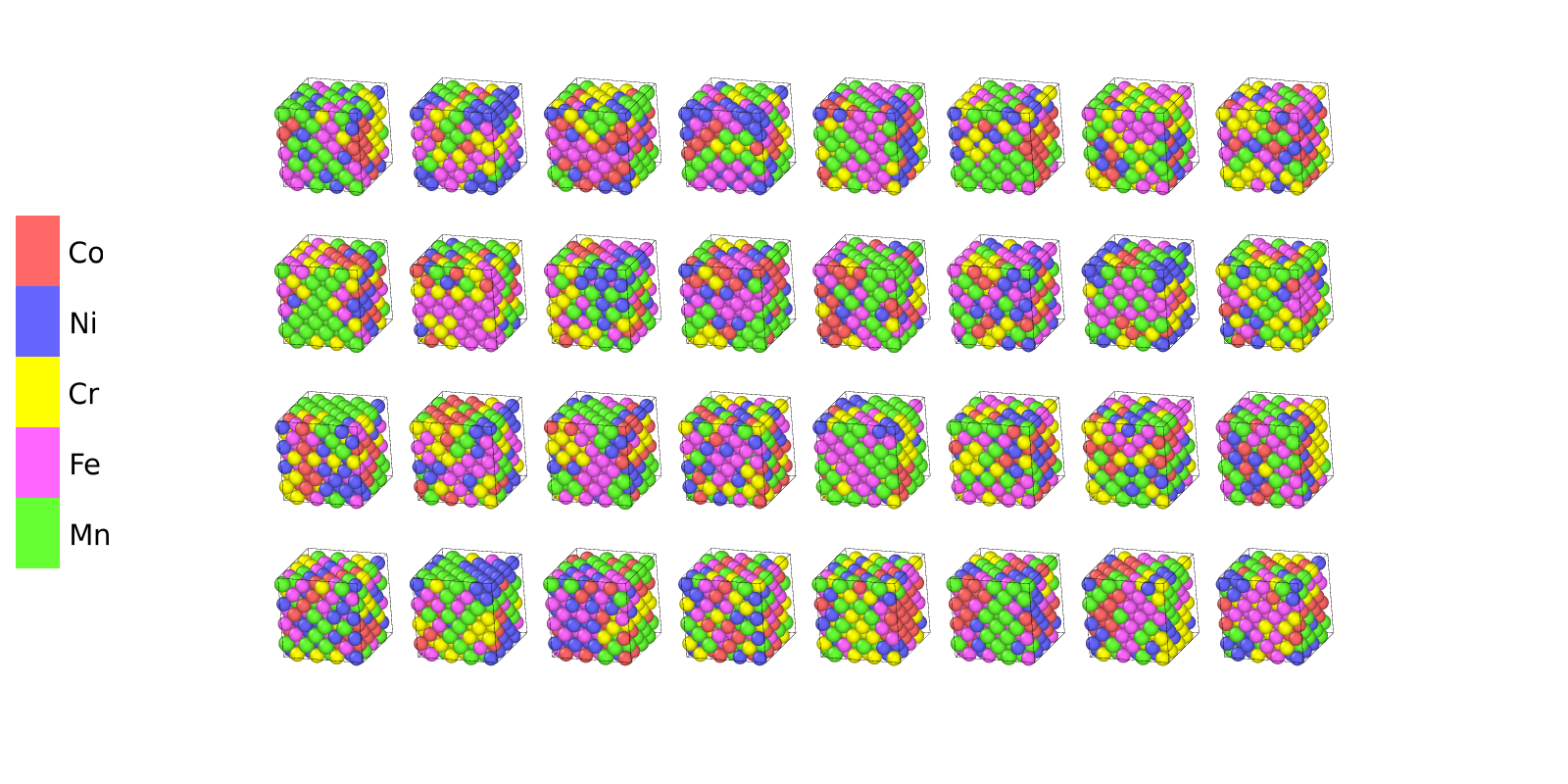}
    \caption{Distribution of SROs in training data for each alloying type pair generated from surrogate models (top) and 32 example configurations (bottom) out of the 500 total.}
    \label{fig:sro-dist-surrogate}
\end{figure}

\begin{figure}[H]
    \centering
    \includegraphics[width=\linewidth]{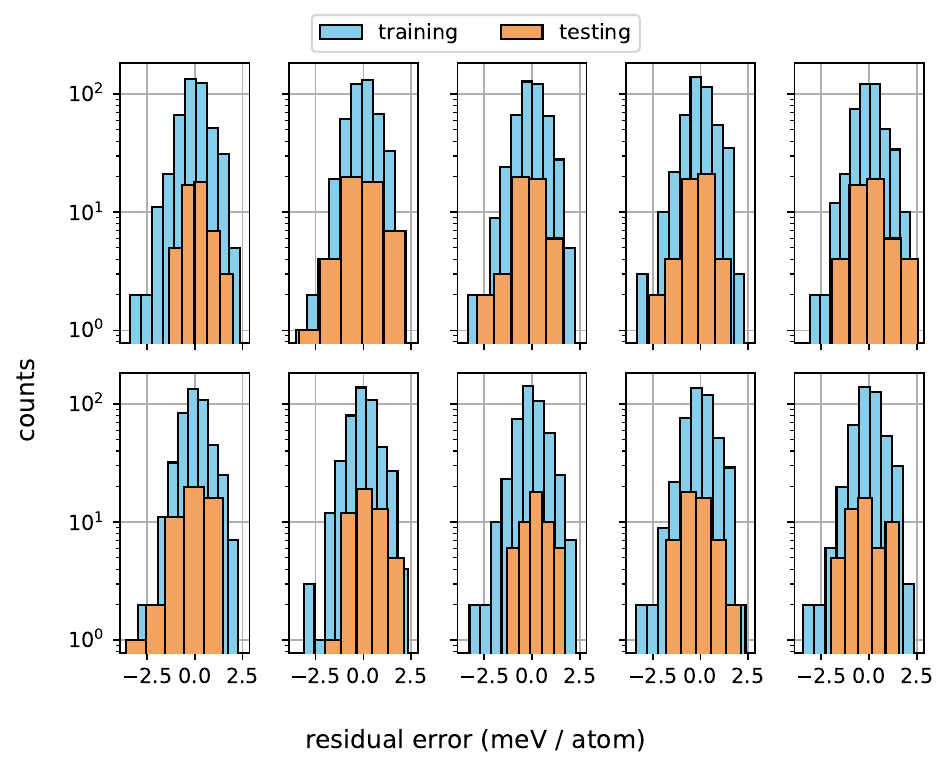}
    \caption{CoNiCrFeMn cross validation residual errors. Each plot represents a different $90\%/10\%$ train-test split.}
    \label{fig:cantor-cross-val}
\end{figure}

Equivalently to the genetic algorithm used for TaW, it is clear that the training set spans a large subset of chemical ordering space. It is additionally clear that the CE model is not over-fitted (since the training and testing residuals are similar), and that we have included a sufficient number of clusters to capture the complexity of the training set (since the residual errors are symmetric about $0$). Additionally, the fitted model is very accurate, with residual errors between $\SI{-3}{meV/atom}$ and $\SI{3}{meV/atom}$ (Figure \ref{fig:cantor-cross-val}).

Note that our genetic algorithm used for TaW is likely to produce a similarly diverse training set for a quinary alloy as well, given an appropriate choice of seeds. However, the number of ordered sub-systems increases quickly with the number of alloying types, making the selection of seeds difficult for a quinary alloy like CoNiCrFeMn. Therefore, it is advantageous to use a data generation technique that is agnostic of an arbitrary choice of initial seeds. Furthermore, using surrogate models like above is more parallelizable, since each equilibration of each surrogate model is independent of the others.

\begin{figure}[H]
    \centering
    \begin{overpic}[width=\linewidth]{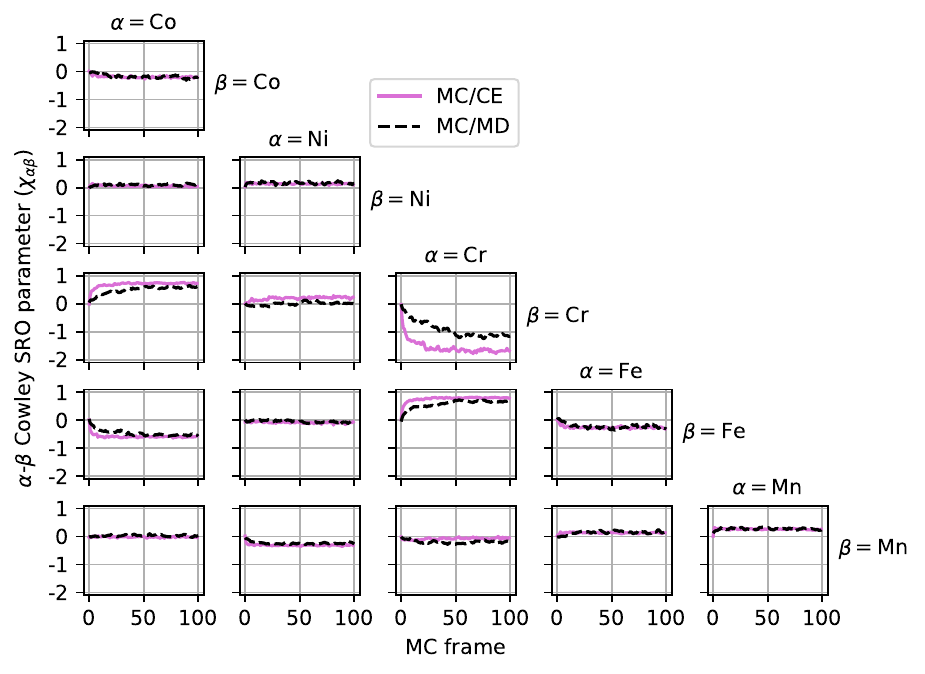}
    \put(60,40){\includegraphics[width=0.4\linewidth]{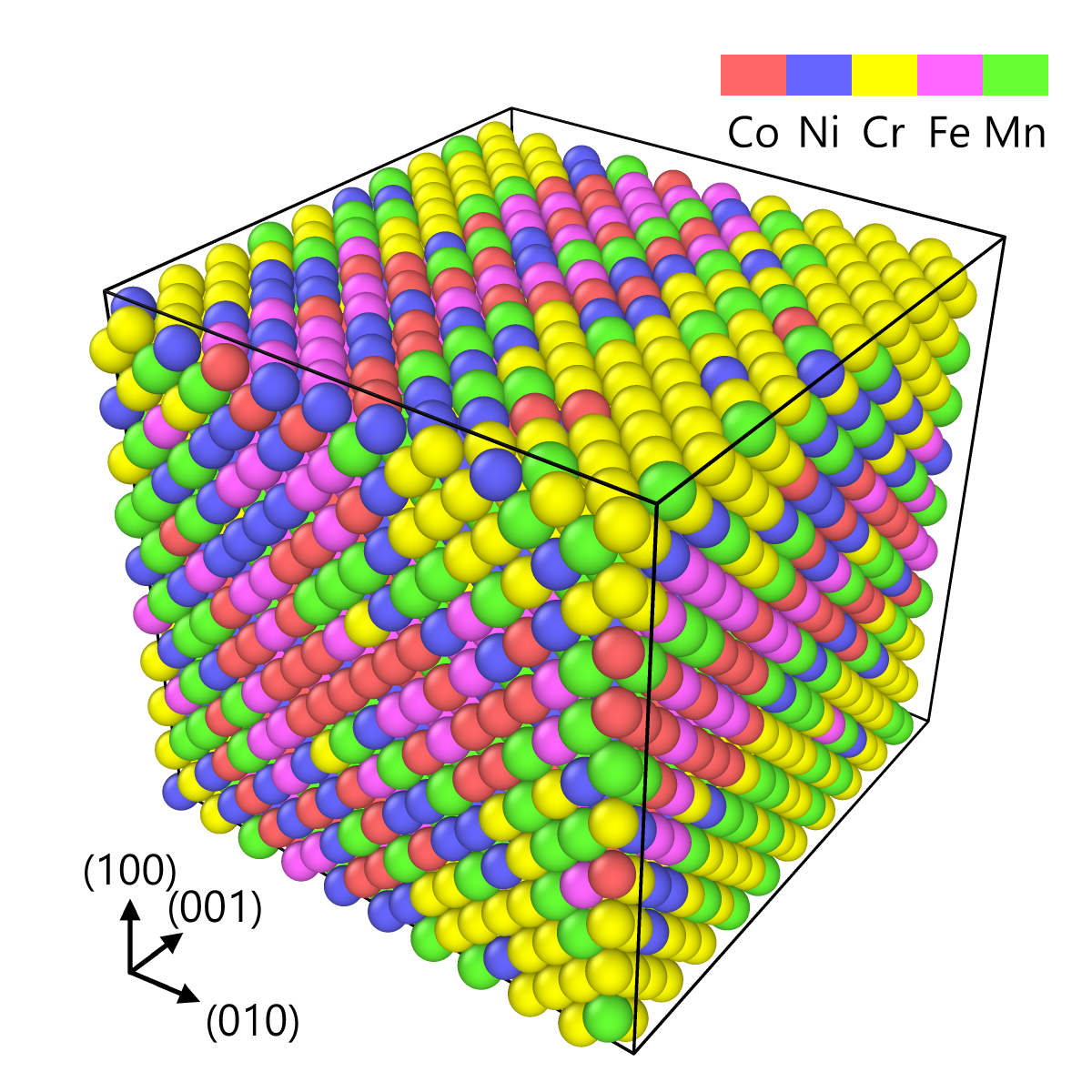}}
  \end{overpic}
    \caption{CoNiCrFeMn Cowley SRO parameters $\chi_{\alpha\beta}$ predicted by Metropolis MC, ran with a CE model and MC/MD with LAMMPS, and a visualization of the final configuration in the MC/CE run. Each MC frame is $10,000$ MC steps.}
    \label{fig:cantor-mc}
\end{figure}

The resulting CE model accurately predicts the full set of Cowley SRO parameters for the CoNiCrFeMn alloy after equilibrating with Metropolis MC at $\SI{600}{K}$ from a fully random solution (Figure \ref{fig:cantor-mc}). The CE/MC routine seems to overestimate the magnitude of the $\text{Cr}$-$\text{Cr}$ SRO parameter $\chi_{\text{Cr}\text{Cr}}$ with respect to the MC/MD routine run with LAMMPS, i.e. the amount of self-segregation of $\text{Cr}$ is overestimated. However, importantly, the $\text{Cr}$-$\text{Cr}$ SRO parameter is predicted to be strongly negative, accurately predicting a strong tendency for $\text{Cr}$ to self-segregate. This is in agreement with experiments performed by Otto et al., which shows the precipitation of a Cr-rich phase at $\SI{500}{^\circ C}$ after sufficient annealing\cite{OTTO201640}. We note that this comparison is a bit strenuous, however, since the MC routines deployed above are on a single-phase fcc lattice, but experimentally the Cr-rich phase is bcc and precipitates mostly at grain boundaries. A more careful CE study would include an analysis of multi-phase systems and/or polycrystals, but this is not trivial, and is therefore saved for future work. We also note that the other SRO parameters for all alloying type pairs are in excellent quantitative agreement with the MC/MD routine results from LAMMPS.
\section{Limitations}

The two primary limitations of the TCE formalism presented in this work are the same as the standard CE formalism, i.e. that the effective Hamiltonian is only mappable to a fixed lattice, and that long-range interactions are difficult to capture.

The former limitation is potentially problematic for systems, for example, with large lattice strain. Additionally, this limitation makes dealing with phase equilibria between different lattice types nontrivial. However, there is work done for computing energetics on mixed lattices, for example the variable-lattice CE developed by Yuge\cite{vce}.

The latter limitation is potentially problematic for systems with long range interactions, such as ionic solids. In principle, this limitation is not difficult to handle, i.e. add more clusters until the energetics are fully captured. However, the number of possible cluster types combinatorically explodes with the chosen cutoff distance, yielding a large number of interaction terms, which can defeat the efficiency of a CE evaluation. Long range effects have indeed been addressed in literature for the standard CE formalism. \cite{PhysRevB.51.11257, oxides, Seko_2014}. A commonly used solution for this problem is simply adding an electrostatic point term, effectively decomposing the system's energy into cluster terms and an electrostatic term:

\begin{equation}
    \mathcal{H}_\text{eff}' = \mathcal{H}_\text{eff} + \frac{1}{\varepsilon_r}E
\end{equation}

where $E$ is simply the electrostatic energy of the system, calculated via a Coulomb potential, and $\varepsilon_r$ is an effective dielectric constant, often computed as a fitting parameter, $\mathcal{H}_\text{eff}$ is the interaction energy from clusters, and $\mathcal{H}_\text{eff}'$ is the total interaction energy. This solution has been thoroughly tested empirically in previous literature\cite{Seko_2014, PhysRevLett.98.266101, C6EE02094A}, and would be a trivial addition to the TCE formalism.
\section{Conclusions}

In this work, we introduce a new CE formalism which evaluates correlation functions entirely using mixed tensor contractions depending on precomputed topology tensors, which generalizes the cluster expansion formalism and its corresponding implementation to any periodic system, including exotic or low-symmetry lattices. This methodology is additionally very vectorizable, allowing for the exploitation of massively parallel architectures like GPUs or TPUs, and eliminating the need to iterate over particular cluster types. We then show that local energy differences are a natural consequence of this formalism, which enable nearly $\mathcal{O}(1)$ energy difference calculations, which are necessary for efficient MC simulations.

We then use this formalism to fit CE models for the $\text{Ta}_{1-x}\text{W}_x$ binary alloy with DFT data and the CoNiCrFeMn alloy with MEAM data. For each system, we respectively compute an enthalpy of mixing curve and the full set of Cowley SRO parameters using their corresponding CE models. Both CE models show excellent agreement with corresponding ground-truth data.
\section{Data Availability}

The package within implementing the formalism developed in this study, \verb|tce-lib|, is available on GitHub\cite{tce_lib}. The resulting simulation data will be made available upon request.
\section{Acknowledgements}

Authors acknowledge support from the U.S. Department of Energy, Office of Basic Energy Sciences, Materials Science and Engineering Division under Award No. DE-SC0022980.

Additionally, this material is based on work supported by the National Science Foundation under Grant Nos. MRI\# 2024205, MRI\# 1725573, and CRI\# 2010270 for allotment of compute time on the Clemson University Palmetto Cluster.
\section{Disclaimer}

Any opinions, findings, and conclusions or recommendations expressed in this material are those of the author(s) and do not necessarily reflect the views of the National Science Foundation.

\bibliography{bibfile.bib}

\end{document}